\documentclass{aa}  

\usepackage{graphicx}
\usepackage{txfonts}
\usepackage{hyperref}

\usepackage{float}

\usepackage[separate-uncertainty=true]{siunitx}
\usepackage{booktabs}

\begin{document}
\renewcommand{\today}{22 July 2026}

   \title{The magnetic field strength of a methane dwarf measured from its radio spectral cutoff}

   \author{T. W. H. Yiu
          \inst{1,2}\fnmsep\thanks{Corresponding email: yiu@astron.nl}
          \and
          H. K. Vedantham\inst{1,2}
          \and
          J. R. Callingham\inst{1,3}
          \and
          T. W. Shimwell\inst{1,4}
          }

   \institute{ASTRON, The Netherlands Institute for Radio Astronomy, Oude Hoogeveensedijk 4, Dwingeloo, 7991 PD, The Netherlands
         \and
             Kapteyn Astronomical Institute, University of Groningen, PO Box 72, 9700 AV, Groningen, The Netherlands
         \and
             Anton Pannekoek Institute for Astronomy, University of Amsterdam, Science Park 904, 1098 XH, Amsterdam, The Netherlands
        \and
             Leiden Observatory, Leiden University, PO Box 9513, 2300 RA Leiden, The Netherlands
             }

   \date{Received 24 November 2025 / Accepted 21 July 2026}
 
  \abstract{Brown dwarfs and gas-giant planets are expected to emit coherent radio emission via the electron cyclotron maser instability (ECMI) mechanism powered by non-thermal electrons in their magnetospheres. Detection of a characteristic spectral cutoff in their cyclotron emission is likely to be the only feasible route to measure their magnetic field strengths. Previous radio measurements of a subset of cold brown dwarfs (spectral type T or methane dwarfs) have not shown such a cutoff at frequencies as high as $\gtrsim 10\,{\rm GHz}$, implying field strengths far higher than canonical expectations from dynamo scaling laws. However, it is also possible that these high-frequency emissions did not originate in the large-scale magnetic field but rather from small high-field-strength loops at the surface, which the scaling laws are not designed to predict. Here we present radio observations of a methane dwarf binary (WISEP J101905.63+652954.2) whose radio emission was first detected in an untargeted low-frequency survey, making it more likely that the emission traces the object's large-scale magnetic field. The object's spectrum shows a clear radio spectral cutoff whose location yields a polar surface field strength of around 126 gauss. Canonical dynamo scaling laws based on energy balance now over-predict the field strength for this object, but the scaling law based on Lorentz-Coriolis force balance (Elsasser number rule) is consistent with the measurement. Our results suggest that multi-frequency (MHz to GHz) monitoring of brown-dwarf electron cyclotron maser emission is essential to separately measure the `mean field' on large scales and `fluctuating' fields on smaller spatial scales in order to determine the dynamo scaling law that most accurately predicts field generation in the metallic hydrogen layer of gas giants and cold brown dwarfs.
  }
   \keywords{Brown dwarfs --
             Dynamo --
             Radio continuum: stars -- 
             Radiation mechanisms: non-thermal
             }

   \maketitle

\section{Introduction}
\label{sec:intro}
Cold brown dwarfs beyond their deuterium-burning phase and gas-giant exoplanets likely have a similar interior structure and magnetic field generation processes in their metallic hydrogen layers \citep{hubbard1997, fortney2010, reiners2010}.
Measurements of the magnetic field strengths in a sample of these objects will allow us to test dynamo models that predict the magnetic field strengths of these objects (e.g. \citealp{chabrier2006, christensen2009, christensen2010}). Current techniques for measuring magnetic field strengths include Zeeman Doppler imaging (ZDI; e.g. \citealp{semel1989, brown1991, donati2006, donati2008, morin2008, morin2010}) and radio observation of coherent emission due to electron cyclotron maser instability (ECMI; e.g. \citealp{wu1979, treumann2006, hallinan2008, hallinan2015, williams2014, kao2016, kao2018}). The former technique becomes extremely challenging due to the lack of detectable non-broadened Zeeman-sensitive spectral lines in dwarfs later than M9, let alone exoplanets. The latter technique hinges on the fact that emission by ECMI is expected to occur at the local cyclotron frequency $\nu_c$ or its second harmonic, which depends solely on the ambient magnetic field strength in the non-relativistic limit \citep{melrose_dulk1982}:
$\nu_c \approx 2.8~B_{\rm gauss}$\,MHz.
Because no observable ECM emission originates under the surface of a brown dwarf (specifically, its ionosphere), this results in an abrupt drop in the radio spectrum at a characteristic cutoff frequency corresponding to the polar field strength on the brown dwarf's surface. Such a spectral cutoff has been observed on Jupiter at around 40\,MHz (corresponding to $B_{\max} \approx 14$\,G) and on other magnetised solar system planets at $<10$\,MHz \citep{zarka_bible2007}. Outside the solar system, this measurement technique has been applied to brown dwarfs \citep{kao2018}.

There exist many dynamo scaling laws that relate the magnetic field strength of a fully convective object (i.e. ultracool dwarfs and gas-giant exoplanets) to the fundamental properties of the object, including core radius $R_c$, core density $\rho$, electrical conductivity $\sigma$, rotation rate $\Omega$, and convected energy flux $q_c$. Here, the term `core' refers to the `dynamo volume', where field generation occurs. In brown dwarfs and gas giants, this corresponds to the metallic hydrogen layer. The laws lead to different predictions of the large-scale surface magnetic field strength. 
For example, some scaling laws assume a balance of Coriolis force and Lorentz force in the dynamo region (e.g. \citealp{stevenson1979}), and thus the quantities $\Omega$ and $\sigma$ become relevant. Other scaling laws, such as \citet{christensen2006}, instead consider the energy flux that is available to balance ohmic dissipation. This leads to a model in which the dynamo saturates for sufficiently fast rotation. In this model, the final magnetic moment depends only on the saturation value set by the available heat energy driving the dynamo.
For a review of these dynamo scaling laws' theoretical foundation and their applications to planets, we refer the reader to \citet{christensen2010}. These scaling laws essentially reproduce the magnetic field strengths of the two gas giants in the Solar System but when extrapolated to exoplanets and brown dwarfs, lead to widely different predictions. As there is currently no definitive radio detection of exoplanets, observing ECM emission from methane dwarfs (i.e. T and Y dwarfs) provides a pragmatic way to test these models since the magnetism and atmospheres of methane dwarfs can be regarded as gas-giant exoplanet analogues (e.g. \citealp{williams2018}, \citealp{manjavacas2021}). Finding the radio spectral cutoff in the ECM emission of a methane dwarf would therefore greatly constrain dynamo models. 

A spectral cutoff has yet to be definitively detected in brown dwarfs. \citet{rose2023} recently detected a T8 brown dwarf between 0.9 and 2.0\,GHz and showed tentative evidence for a spectral cutoff, although more sensitive observations are necessary to determine whether it is indeed a cutoff or is instead due to spectral evolution of the brown dwarf's radio luminosity. One reason for the lack of cutoff detections is that radio detections of methane dwarfs are generally rare; dedicated targeted searches so far have only yielded seven radio-emitting methane dwarfs \citep{route2012, route2016, kao2016, rose2023}. Using the Karl G. Jansky (Expanded) Very Large Array (VLA), \citet{kao2018} followed up five late L and T dwarfs at X-band (8--12\,GHz) previously detected by VLA or the Arecibo telescope, but a spectral cutoff was absent in the radio spectra of their targets.
This is unexpected since the cutoff frequencies of these targets should be below X-band according to the widely used heat-flux scaling law by \citet{christensen2009} and the brown dwarf evolutionary model by \citet{baraffe2003}. This suggests that other dynamo scaling laws must be considered and/or that our interpretation of the magnetic structure in which the radio emission originates must be reconsidered.

An alternate explanation for the lack of a cutoff detection is the selection bias from the VLA/Arecibo-discovered samples; these targets were previously identified as radio-bright in gigahertz-frequency searches, which were more likely to discover objects that have much stronger fields, perhaps in localised small-scale structures. 
It is therefore plausible that they detected objects with radio emission originating from the strong magnetic fields in small-scale magnetic loops, whereas the dynamo models only predict the average magnetic field strength that is dominated by the large-scale field in these objects. 
Such intense smaller-scale fields are absent in gas giants but are present on the Sun, which has a large-scale field of only 1 gauss (G) but can have kG-level fields in small coronal loops. It is currently unclear whether cold brown dwarfs harbour such strong small-scale fields, although Zeeman Doppler Imaging (ZDI) of fully convective late M dwarfs suggests that these objects either have strong dipole-dominated fields or small-scale-dominated multipolar fields \citep{morin2010}. 

Distinguishing between emission originating from the large-scale `mean field' and from smaller-scale magnetic structures requires multi-frequency monitoring of brown dwarfs across the MHz-to-GHz regime. If the low-frequency ECM emission traces the large-scale field while higher-frequency emission originates from localised high-field-strength loops, then the low-frequency component should disappear above some characteristic cutoff frequency, while a distinct higher-frequency component emerges with a different rotational modulation profile. Furthermore, if the smaller-scale magnetic structures correspond to a fluctuating dynamo component, then the morphology of the high-frequency radio light curve may evolve with time as individual magnetic loops form and dissipate. Multi-frequency monitoring therefore provides a means to observationally separate large-scale and small-scale magnetic field components in ultracool dwarfs.

Detecting brown dwarfs at frequencies much lower than $\sim1$\,GHz is therefore necessary to obtain a complete picture of the morphology and strength of brown dwarf magnetospheres. In fact, low-frequency radio surveys have the potential to unveil a new population of brown dwarfs that host weaker ($<$kG) magnetic fields plus the ones that are too cold to have any detectable emission in canonical infrared sky surveys. Using the Low-Frequency Array (LOFAR; \citealp{lofar}) High Band Antennas (HBAs) observing at 120--168\,MHz, \citet{harish_elegast2020} illustrated such potential with the discovery of the first radio-selected brown dwarf, i.e. the brown dwarf was not identified at any other wavelength prior to its radio detection.
Separately, \citet{harish_j1019} also discovered a new radio-loud brown dwarf system called WISEP J101905.63+652954.2 ($=$~2MASS J10190575+6529526~$=$~WISEA J101905.61+652954.0; hereinafter J1019+65) using LOFAR. J1019+65 is the object of study for this paper. It consists of two brown dwarfs in a binary with spectral types T$5.5 \pm 0.5$ and T$7.0 \pm 0.5$ \citep{harish_j1019}, and is at a distance of $\num{23.3 \pm 1.0}$\,pc \citep{kirkpatrick2019}. The two brown dwarfs have a projected separation of $423.0 \pm 1.6$\,mas, which corresponds to $\num{9.9 \pm 0.4}$\,AU \citep{harish_j1019}. J1019+65 showed circularly polarised radio pulsations at 144\,MHz in data from the LOFAR Two-metre Sky Survey (LoTSS; \citealp{lotss_dr2}).

J1019+65 is a well-suited system to test dynamo models for multiple reasons.
Firstly, since LoTSS is an untargeted all-sky survey, the radio detection of J1019+65 does not suffer from selection bias, which is inherent to targeted observations. Thus, by using J1019+65 as an archetype, we can better generalise our conclusions to the population of radio-loud brown dwarfs.
Secondly, unlike the VLA- and Arecibo-detected brown dwarfs, J1019+65 does not suffer from the selection bias towards a kG-level magnetic field. Indeed, J1019+65 was detected at a frequency much smaller than the predicted cutoff in its large-scale field according to the heat-flux scaling law of \citet{christensen2009}.
Thirdly, unlike isolated brown dwarfs, the binary nature of J1019+65 allows for an independent mass constraint using infrared observations \citep{harish_j1019}. The mass of a brown dwarf is a vital input parameter for existing fully convective dynamo models. Therefore, detecting a spectral cutoff in a brown dwarf with known mass will provide the tightest constraint on these dynamo models.

In this paper, we present our latest observational campaign of J1019+65 that utilised three different radio telescopes: LOFAR, the upgraded Giant Metrewave Radio Telescope (GMRT), and VLA. The campaign comprises more than 70 hours of radio observations and spans almost 8 years (from 2017-03-08 to 2025-01-07), making it one of the longest and most intensive radio monitoring campaigns of a brown dwarf system to date.
The observations were aimed at detecting radio emission from J1019+65 in order to find the cyclotron-frequency cutoff in the radio spectrum of J1019+65.
The paper is structured as follows. In Sect.~\ref{sec:obs}, we present the follow-up observations of J1019+65 by the three radio telescopes, the data reduction procedures and the resulting flux density measurements showing a spectral cutoff. We discuss our results, especially the radio cutoff detection and its implications for dynamo scaling laws in Sect.~\ref{sec:discussion} and conclude in Sect.~\ref{sec:conclusion}.

\section{Observations and data reduction}
\label{sec:obs}

Since the discovery of J1019+65's radio emission in LoTSS DR2, we have acquired 15 new radio observations to follow up on J1019+65 using three radio telescopes: LOFAR, GMRT, and VLA as part of projects \texttt{LC15\_019} \& \texttt{LT16\_013}, \texttt{ddtC252} \& \texttt{47\_039}, and \texttt{22B\_332} respectively. More details on each follow-up radio observation of J1019+65 -- such as the observing frequency range, start time, and length of observation -- are listed in Table~\ref{tab:observations} in the Appendix.

\subsection{Why GMRT and VLA observations?}
\label{sec:why}
In the absence of a clear cutoff in the spectrum of J1019+65 in the LOFAR 120--168\,MHz band \citep{harish_j1019, yiu2025}, we acquired observations of the target with the GMRT telescope in band-3 and -4 (300--950\,MHz) and with the VLA in L-band (1--2\,GHz).
We targeted higher frequencies with GMRT and VLA to search for emission above the LOFAR band.

An additional advantage of the GMRT is that contemporaneous LOFAR observations can be scheduled. Contemporaneous observations are necessary to check if any cutoff detection is an artefact of the inherent variability in J1019+65's radio emission. As seen in Table~\ref{tab:observations}, two of the four GMRT observations (2023-01-24 and 2023-03-08) had contemporaneous LOFAR data.

\subsection{Data reduction procedure}
\label{sec:procedure}
The procedures for calibrating the datasets from LOFAR, GMRT, and VLA are described in Sects.~\ref{sec:obs-lofar},~\ref{sec:obs-gmrt}, and~\ref{sec:obs-vla} respectively. The post-calibration analysis for all datasets follows these procedures:
For each calibrated dataset, we used \texttt{wsclean} \citep{wsclean2014} to create Stokes V and I dirty images synthesised over the full bandwidth (see Table~\ref{tab:observations} in the Appendix for the frequency band of each observation) and phase-shifted to the proper-motion-corrected sky location of J1019+65 according to IR astrometric information \citep{kirkpatrick2019}.
Moreover, we created a dynamic spectrum in Stokes V for each observation from the phase-shifted visibilities to search for time- or frequency-dependent emission that may not be visible in synthesised images.
This was done for each observation by averaging the visibilities (from the phase-shifted calibrated dataset) across baselines and then binning the data in both time and frequency.

\subsubsection{LOFAR observations}
\label{sec:obs-lofar}
As seen in Table~\ref{tab:observations}, there are a total of 10 LOFAR follow-up epochs after the original LoTSS 2017 epoch.
Except for the first LOFAR follow-up observation in 2021 (project code: \texttt{LC15\_019}), which has the same exposure time (8 hours) as the LoTSS field, all other LOFAR observations (project code: \texttt{LT16\_013}) lasted 4 hours each to ensure that we covered the full 3-hour rotation (with a 2$\sigma$ uncertainty range of 2.6--3.8 hours; \citealp{harish_j1019}) of J1019+65 in each observation. These are the same datasets as those presented by \citet{yiu2025}. In that article, the time variability in the light curve of J1019+65 using LOFAR was explored, whereas here we are interested in its spectral evolution.

The LOFAR data reduction followed that in \citet{yiu2025} but we summarise it here for completeness. The dataset for each epoch was first calibrated for direction-independent effects using the LOFAR Initial Calibration (LINC) pipeline\footnote{\url{https://git.astron.nl/RD/LINC}} (see \citealp{linc_degasperin2019} for details).
Subsequently, each dataset underwent direction-dependent self-calibration and imaging using the \texttt{DDFacet} pipeline\footnote{\url{https://github.com/mhardcastle/ddf-pipeline}} \citep{tasse2021,lotss_dr2}.
We then followed the extraction and self-calibration procedure by \citet{van-weeren2021}.
We made Stokes V and Stokes I images of J1019+65 for each LOFAR epoch from the pipeline-processed datasets using \texttt{wsclean} with Briggs weighting \citep{briggs1995}; a robustness parameter of $0.0$ was used for Stokes V to achieve higher sensitivity, while a more negative robustness parameter of $-0.5$ was used for Stokes I to suppress sidelobes from bright-source contamination (as there are many more bright sources in the Stokes I sky compared to Stokes V).
In addition to full-band imaging, we produced `sub-band' images by setting \texttt{channels-out=3} in \texttt{wsclean}, thereby dividing the observing band into three equal-width frequency intervals (i.e. 120--136\,MHz, 136--152\,MHz, and 152--168\,MHz).
For each of the three frequency intervals, we produced one image per observing epoch. The images corresponding to the same frequency interval were then combined across epochs using inverse-variance weighting to improve sensitivity to persistent emission from J1019+65.

\subsubsection{GMRT observations}
\label{sec:obs-gmrt}
We acquired two observations of J1019+65 using GMRT with the band-4 (550--950\,MHz) and band-3 (300--500\,MHz) receivers on 2023-01-24 and 2023-03-08 respectively. These two GMRT observations overlapped partially with the two LOFAR observations of the same date (see Table~\ref{tab:observations}). Since LOFAR is a phased array, its sensitivity drops significantly when observing below $\approx 45\degr$ elevation. Therefore, it is difficult to schedule a full overlap between the observations carried out by LOFAR and GMRT while still achieving sufficient sensitivity to detect the pulse with LOFAR. We chose to prioritise the latter.

For the GMRT band-4 observation, we used the CAsa Pipeline-cum-Toolkit for Upgraded Giant Metrewave Radio Telescope data REduction (\texttt{CAPTURE}) pipeline \citep{capture-kale2021} to reduce the data and imaged the target field in Stokes V using \texttt{wsclean} with Briggs weighting (robustness parameter = 0.0).
However, unlike the \texttt{DDF-pipeline} for LOFAR data processing, the \texttt{CAPTURE} pipeline does not perform direction-dependent self-calibrations.
Therefore, the \texttt{CAPTURE} pipeline product for the GMRT band-3 observation was dynamic-range limited due to a bright in-field source (see Sect.~\ref{sec:leakage-jy-source}). Instead, we manually reduced the GMRT band-3 data using the Common Astronomy Software Applications (\texttt{CASA}; \citealp{casa2022}) software, following standard flagging and calibration procedures. This includes multiple rounds of self-calibration. 
Of all the radio data in this campaign, the band-3 epoch is by far the most challenging to reduce. In Sect.~\ref{sec:leakage-jy-source}, we discuss the quantitative effects of the bright source on each band, and whether the fidelity of the synthesised images is worsened due to its presence.

In addition, in 2025 we observed J1019+65 with GMRT in band-3 again twice, without contemporaneous LOFAR exposures. The data were manually reduced using CASA.

\subsubsection{VLA L-band observation}
\label{sec:obs-vla}

For reasons discussed in Sect.~\ref{sec:why}, we observed J1019+65 in L-band on 2023-04-22. The VLA data went through the standard VLA calibration pipeline\footnote{\url{https://science.nrao.edu/facilities/vla/data-processing/pipeline}} (\texttt{CASA} version 6.5.4-9; pipeline version 2023.1.0.125), and the pipeline-processed data were used to image our target using \texttt{wsclean} with natural weighting to achieve the lowest RMS noise in the Stokes V image. We used 3C 286 as the flux density calibrator and J1056+7011 as the complex gain (i.e. secondary) calibrator.

\subsection{Leakage from a nearby Jy source}
\label{sec:leakage-jy-source}
Approximately 23\arcmin~away from J1019+65, there exists a bright extended radio source named 4C 66.10, with a Stokes I flux density of 1.1\,Jy at 1.4\,GHz \citep{condon1998} and 2.2\,Jy at 144\,MHz \citep{lotss_dr2}.
Since LOFAR, GMRT, and VLA can all easily reach sub-mJy sensitivity in typical observations, the presence of this bright source can make the synthesised images dynamic-range limited instead of thermal-noise limited.
The dynamic-range limitation is mitigated in Stokes V images because 4C 66.10, being a synchrotron source, should have negligible levels of circularly polarised emission. However, for such a bright source, even a small level of leakage of total intensity into circular polarisation will produce detectable image artefacts (e.g. the LoTSS Stokes V leakage is measured to be around 0.06\%; \citealp{v-lotss}). Moreover, the leakage flux is due to calibration errors and does not follow the instrumental point spread function, meaning that the leaked source cannot be cleaned/deconvolved during imaging.

In the following, we briefly discuss the quantitative effect of 4C 66.10 on the image fidelity for each frequency band used in the campaign.
\begin{itemize}
    \item LOFAR: Owing to direction-dependent self-calibrations in the \texttt{DDF-pipeline}, we were able to subtract the bright source (and its sidelobes) from the LOFAR datasets. We note that the LOFAR Stokes I images still have an RMS noise twice as large as the Stokes V images; in Stokes I the residual sidelobes of the bright source could not be perfectly subtracted due to small calibration imperfections, whereas in Stokes V the source is much dimmer and thus the residual sidelobes are negligible.
    \item GMRT: For the band-4 observation, we were able to reach an RMS noise close to thermal noise since the primary beam (i.e. the full width at half maximum) is smaller at a higher observing frequency. However, for the band-3 observations, the Stokes V images are limited by dynamic range, yielding RMS noise levels approximately 3--8 times\footnote{The range arises because the 2025-01-07 observation yielded better sensitivity than the other band-3 observations owing to its longer exposure time (see Table~\ref{tab:observations}) and better ionospheric conditions.} the thermal-noise level. We choose to report the best noise level per band out of multiple observations in the following section.
    \item VLA: The data quality of the L-band observation was not affected by the bright source. This is because of (a) the smaller beam size of the VLA, meaning that 4C 66.10 is outside the field; and (b) 4C 66.10 is not as bright at higher frequencies. Therefore, we were able to achieve thermal noise in the Stokes V image for the VLA observation with ease.
\end{itemize}

\begin{figure*}
    \centering
    \resizebox{\hsize}{!}{\includegraphics[width=1.0\textwidth]{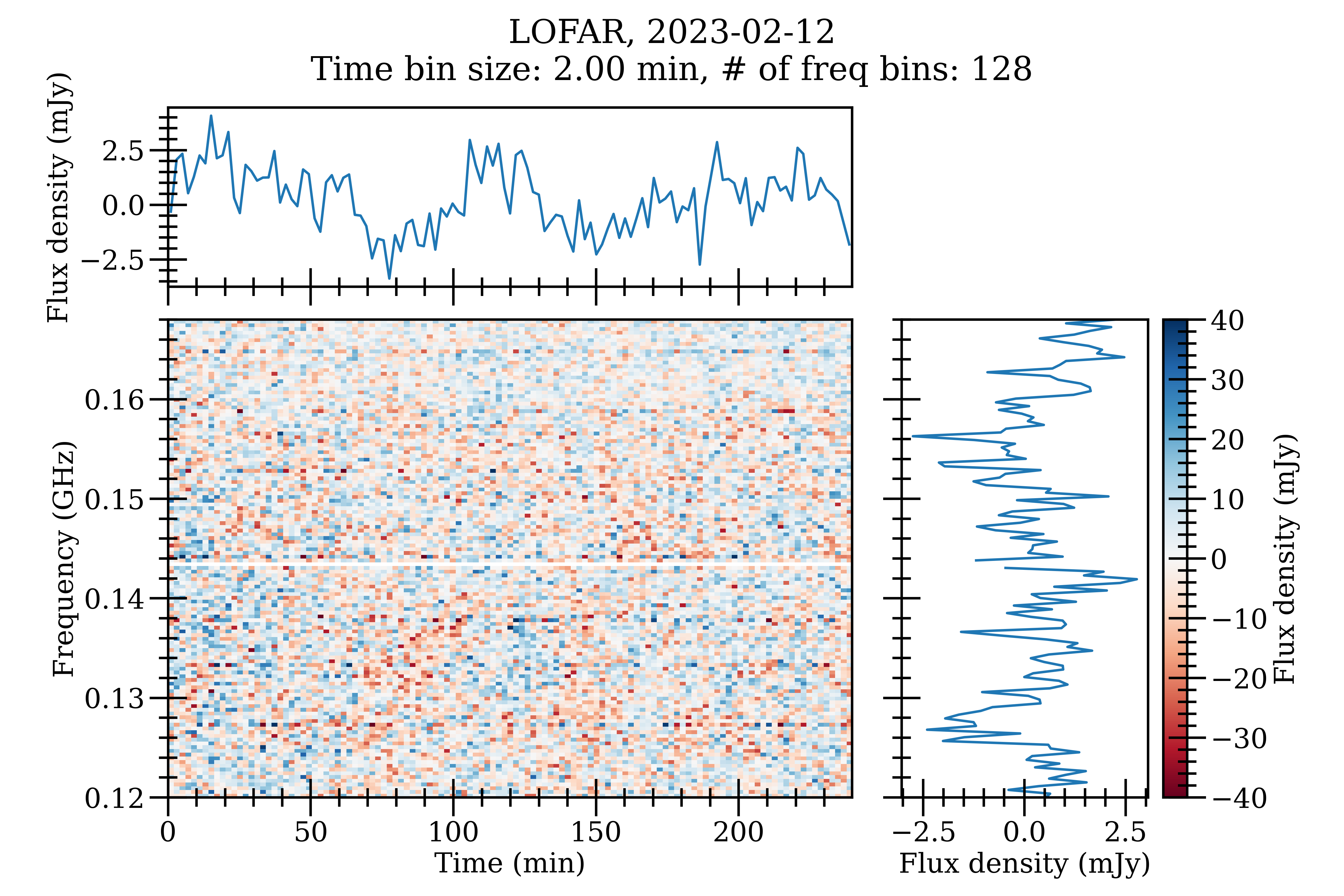}}
    \caption{Dynamic spectrum from one of our LOFAR observations (on 2023-02-12), binned into time bins of 2 minutes and 128 frequency bins. This dynamic spectrum shows no coherent emission from J1019+65; only noise and radio frequency interference (RFI) are present.}
    \label{fig:dynspec}
\end{figure*}

\begin{figure*}
    \centering
    \resizebox{\hsize}{!}{\includegraphics[width=1.0\textwidth]{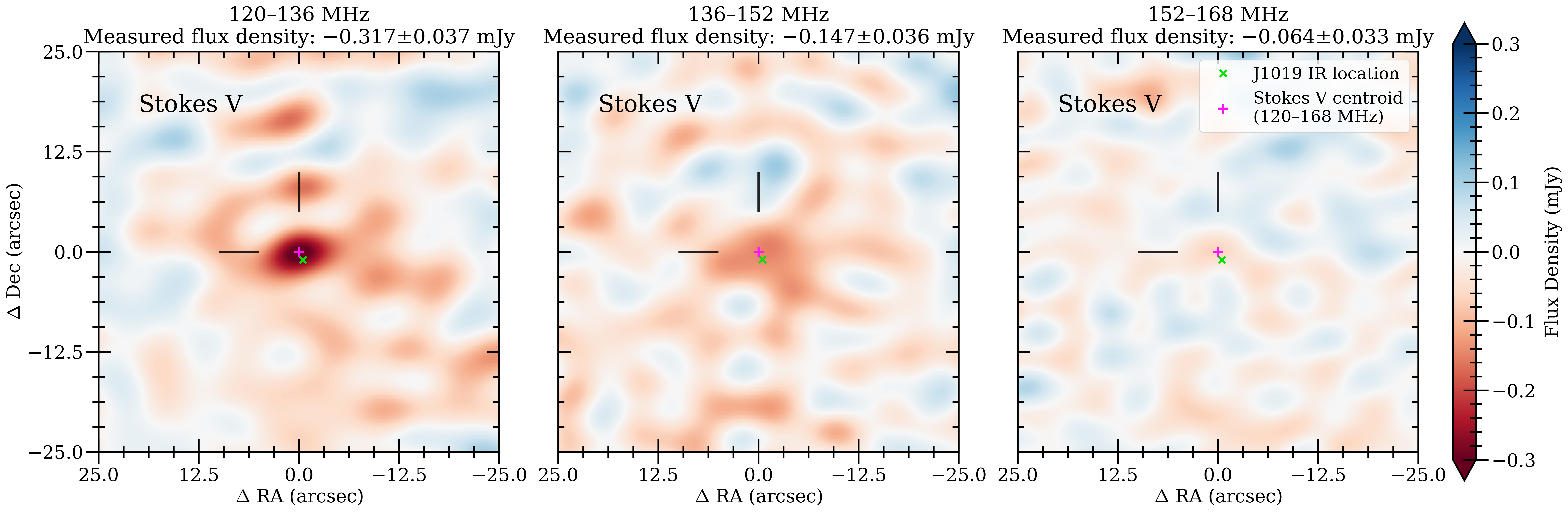}}
    \caption{Inverse-variance-weighted average Stokes V images of J1019+65 from all LOFAR data (2017--2023), centred at the radio centroid (marked in magenta). The flux density of the radio centroid (peak pixel) is $-0.32 \pm 0.04\,$mJy at 120--136\,MHz, $-0.15 \pm 0.04\,$mJy at 136--152\,MHz, and $-0.06 \pm 0.03\,$mJy at 152--168\,MHz. The proper-motion-corrected location of J1019+65 derived from infrared observation is marked by a lime cross, approximately 0.5\arcsec~away from the radio centroid.}
    \label{fig:stacked-image-v-all}
\end{figure*}

\begin{table}[]
\caption{Flux density measurements of J1019+65 for each frequency band in our observational campaign.}
\begin{tabular}{c|r}
$\nu_{\rm obs}$ & $|S_V|$ \\ \midrule
120--168\,MHz  & $160 \pm 20\,\mu$Jy          \\
300--500\,MHz  & $<16\,\mu$Jy           \\
550--950\,MHz  & $<11\,\mu$Jy           \\
1--2\,GHz      & $<5\,\mu$Jy                      
\end{tabular}
\tablefoot{$\nu_{\rm obs}$ and $|S_V|$ are the observing frequency band and absolute Stokes V flux density respectively. For non-detections (shown as upper limits), we report the $1\sigma$ RMS noise of the image. Since J1019+65 was observed in band-3 (300--500\,MHz) three times, we report the lowest RMS noise among three epochs.}
\label{tab:flux}
\end{table}

\subsection{Main results}
\label{sec:results}
The time- and frequency-integrated emission of J1019+65 has a Stokes V flux density of $\approx -0.16 \pm 0.02\,$mJy in the LOFAR band \citep{yiu2025}. 
Although the same level of flux density would easily be detectable in the GMRT (band-3 and band-4) and VLA L-band, the source was undetected in these data in either Stokes V or Stokes I.
The noise levels in the Stokes V images were $\sigma_{\rm V} \approx 16, 11, 5\,\mu$Jy for the GMRT band-3, GMRT band-4, and VLA L-band observations respectively (see Table~\ref{tab:flux}).
Therefore, no time- or frequency-integrated emission is detected in the GMRT and VLA data, indicating a strong decline in emission toward higher frequencies between 144 and 400\,MHz.
This trend is further illustrated in Fig.~\ref{fig:stacked-image-v-all}, which shows the inverse-variance-weighted average Stokes V images of J1019+65 in three frequency intervals across the LOFAR band. The flux density measured in the inverse-variance-weighted image is $-0.32 \pm 0.04$\,mJy at 120--136\,MHz, and it starts to decrease (in magnitude) to $-0.15 \pm 0.04$\,mJy at 136--152\,MHz, until it fades to $-0.06 \pm 0.03$\,mJy at 152--168\,MHz, showing a sharp decrease in flux density across the LOFAR sub-bands.

We also constructed the LOFAR-band dynamic spectra for each observation to glean more information on the spectro-temporal evolution. An example dynamic spectrum from one of our observations is plotted in Fig.~\ref{fig:dynspec}. The dynamic spectra show only noise and residual radio frequency interference (RFI), demonstrating that the sensitivity is insufficient to reveal useful spectral information beyond what is shown in Fig.~\ref{fig:stacked-image-v-all}.

\begin{figure}
    \centering
    \resizebox{\hsize}{!}{\includegraphics[width=1.0\textwidth]{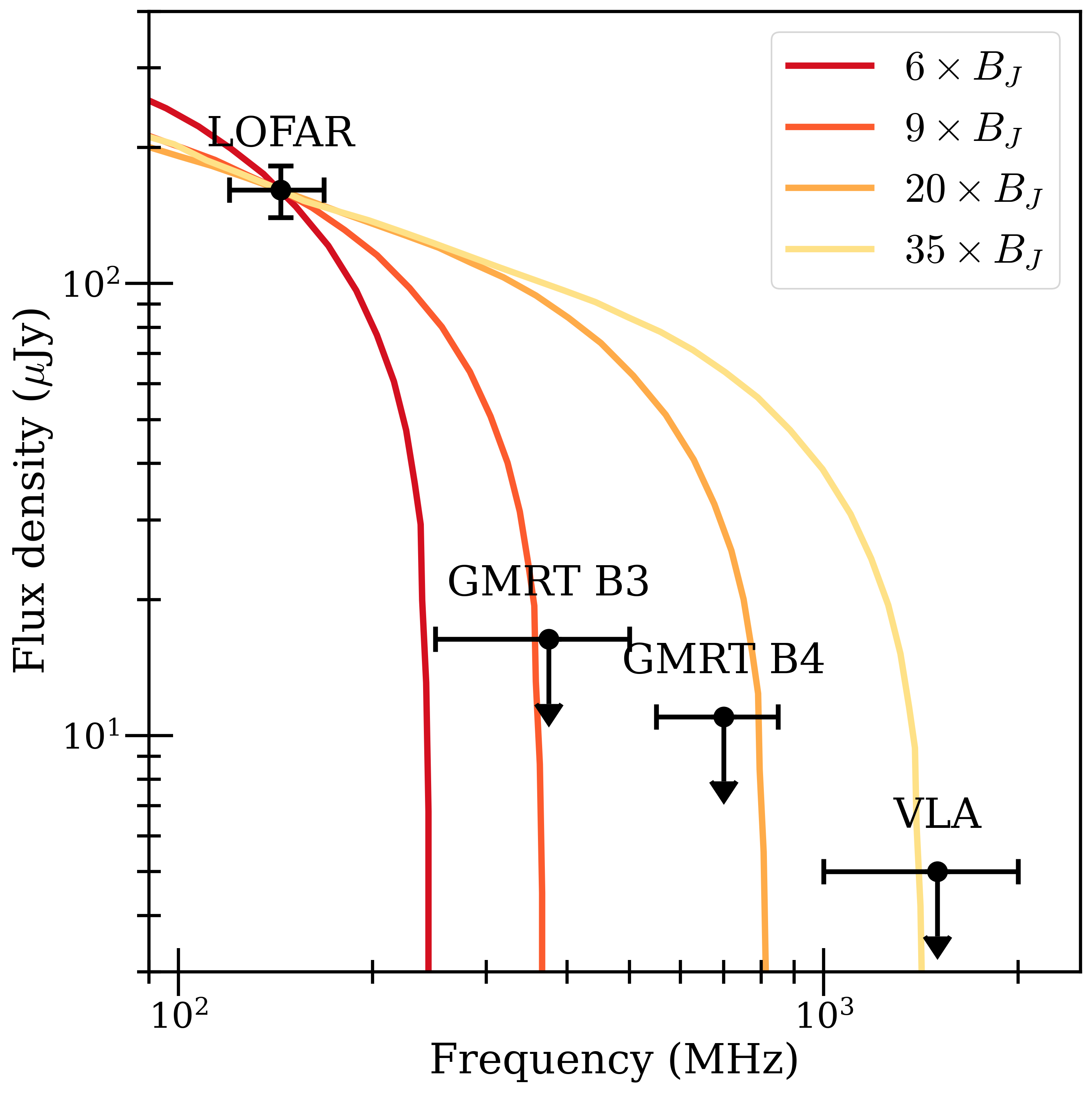}}
    \caption{Radio spectrum of J1019+65 in comparison to the Jovian spectra of different peak magnetic field strengths ($B_{\max} = 6 \times$, $9 \times$, $20 \times$, and $35\times B_J$, where $B_J \approx 14\,$G is the peak Jovian magnetic field strength), corresponding to different characteristic cutoff frequencies (recall ECM emission occurs at $\nu_c \approx 2.8~B_{\rm gauss}$\,MHz). Each spectrum is upscaled in flux density to match the LOFAR-detected value of J1019+65's quiescent emission. The $1\,\sigma$ upper limits obtained from the GMRT band-3/band-4 (B3/B4) and VLA L-band observations are plotted to help constrain J1019+65's spectrum. The original Jovian spectrum (where $B_{\max} = 1 \times B_J$; not shown in this figure) corresponds to the non-burst component of its auroral radio emission and is adapted from \citet{zarka1992,zarka2004,zarka_bible2007}.}
    \label{fig:j1019-spectrum}
\end{figure}

\section{Discussion}
\label{sec:discussion}

\subsection{Spectral cutoff}
\label{sec:discussion-cutoff}
The non-detections by GMRT/VLA and the sharp decrease in flux density from 120 to 168\,MHz suggest that the emission is beginning to cut off in the LOFAR band.
However, due to the finite thermal noise in our images, the spectral evolution of the flux density could appear as a hard cutoff. To determine the surface magnetic field strength of J1019+65, we must distinguish between a hard spectral cutoff and spectral evolution of the flux density due to different emission components as seen in Jupiter's radio spectrum \citep{zarka1998}.
Because the spectral shape of an ECM emitter cannot be predicted from first principles, we make the pragmatic choice of using the Jovian spectrum as a template to represent the plausible spectral variations around the putative cutoff frequency in J1019+65.
In Fig.~\ref{fig:j1019-spectrum} we compare our flux density constraints with the scaled Jovian non-burst auroral radio spectrum from \citet{zarka_bible2007}.
Here, the different coloured lines represent the Jovian spectrum scaled in frequency and flux density. The frequency scaling corresponds to different values of the polar surface magnetic field strength $B_{\max}$ expressed in units of $B_J$, the Jovian value. The spectral scaling, by a factor of $\sim 10^4$, is chosen such that the scaled curves always pass through the LOFAR detection.
Based on Fig.~\ref{fig:j1019-spectrum}, we place a constraint of $B_{\max} \lesssim 9 B_J = 126\,{\rm G}$ on the polar surface field strength of the emitter in J1019+65.

\subsection{Implications for dynamo scaling laws}
\label{sec:discussion-scaling_laws}

\begin{table}[]
\caption{Scaling laws proposed by \citet{christensen2010}.}
\begin{tabular}{l|ll}
\# & Rule & Author \\ \midrule
3  & $B^2 \propto \rho \Omega \sigma^{-1}$                  &    \citet{stevenson1979}            \\
5  & $B^2 \propto \rho \Omega R_c^{5/3} q_c^{1/3}$          &    \citet{curtis1986}            \\
8  & $B^2 \propto \rho \Omega^{1/2} R_c^{3/2} q_c^{1/2}$    &    \citet{starchenko2002}            \\
9  & $B^2 \propto \rho R_c^{4/3} q_c^{2/3}$                 &    \citet{christensen2006}           
\end{tabular}
\label{tab:scaling_laws}
\end{table}

We now compare the magnetic field strength of J1019+65 (derived from the spectral cutoff) to the predicted value of the magnetic energy density $B^2/2\mu_0$ in accordance with the different scaling laws listed in Table~\ref{tab:scaling_laws}. The details of these scaling rules were described by \citet{christensen2010}, but here we briefly mention their respective physical motivations for clarity. Rule \#3 is the so-called `Elsasser number rule', where the strength of the magnetic field is set by the balance between the Lorentz force and the Coriolis force. Rules \#5 and \#8 are derived from the characteristic velocity based on force-balance arguments: mixing-length theory (assuming a characteristic length scale on which convective mixing of momentum and entropy occurs) and MAC balance (from magnetic, Archimedean, and Coriolis forces) respectively. Rule \#9 is unique compared to other scaling laws in that the magnetic field value does not depend on the rotation rate. It posits that the magnetic field's strength is determined by the available convected energy flux that is driving the dynamo.

For each scaling law, we determined the constant of proportionality using Jupiter and Saturn. We chose Jupiter and Saturn, as opposed to other magnetised solar system planets, because they are thought to generate their magnetic fields in fully convective metallic hydrogen layers similar to those in brown dwarfs \citep{saumon2004}. We assumed the following values for Jupiter and Saturn (in SI units) respectively: rotation rates of $\Omega = \num{1.759e-4}$ and $\num{1.653e-4}$, an electrical conductivity of $\sigma = \num{2e6}$ for both\footnote{We assume the same $\sigma$ for brown dwarfs as well, since all these objects generate their magnetic field via their metallic hydrogen layer.}, core radii of $R_c = 0.83 R_J = \num{5.803e7}$ and $0.40 R_S = \num{2.329e7}$ \citep{christensen2009, christensen2010}, core densities of $\rho = \num{1.326e3}$ and $\num{6.871e2}$, and convected energy fluxes of $q_c = 7.485$ and $2.84$ \citep{li2018, wang2024}. Here, the `core radius' is the radius of the metallic hydrogen layer.

\begin{figure*}
    \centering
    \resizebox{\hsize}{!}{\includegraphics[width=1.0\textwidth]{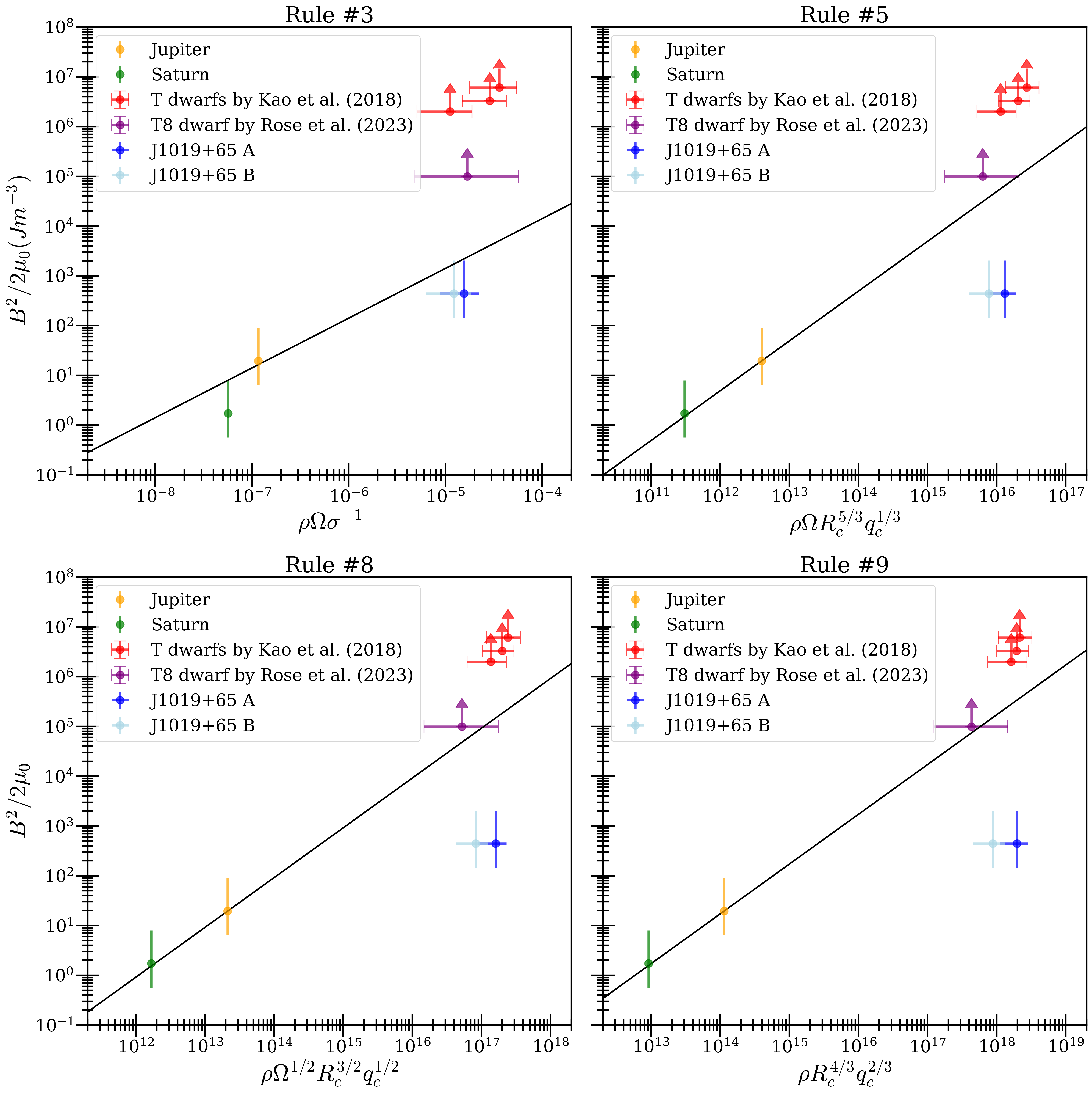}}
    \caption{Magnetic energy density of fully convective objects versus a function of the object's fundamental properties. Each plot shows the magnetic energy density predicted by the corresponding scaling law (black line). All quantities are in SI units. The black line is fitted using Jupiter and Saturn as data points. Since J1019+65 is a binary and we do not know which T dwarf in the binary is the radio emitter, we plotted both T dwarfs (heavier T5.5 component in blue, lighter T7 component in light-blue). The T dwarf detections by \citet{kao2018} and the T8 dwarf by \citet{rose2023} are shown as lower limits (see discussion in Sect.~\ref{sec:discussion-comparison}).}
    \label{fig:scaling_laws_plot}
\end{figure*}

\subsection{Comparison with previous methane dwarf measurements}
\label{sec:discussion-comparison}

To compare our target with other radio-loud T dwarfs in the literature, we also plotted four other T dwarfs in Fig.~\ref{fig:scaling_laws_plot}. These include three T dwarfs whose radio ECM emissions were detected at 8--18\,GHz by \citet{kao2018} using VLA, and one T8 dwarf detected at 0.9--2.0\,GHz by \citet{rose2023} using the Australian Telescope Compact Array and MeerKAT. These are 2MASS 10475385+2124234 (T6.5 dwarf), SIMP J01365662+0933473 (T2.5 dwarf), 2MASS J12373919+6526148 (T6.5 dwarf), and WISE J062309.94$-$045624.6 (T8 dwarf). More details about these T dwarfs (such as mass and rotation rate) and their radio emission were described by \citet{kao2018} and \citet{rose2023}.

The radio emissions of the T dwarfs observed by \citet{kao2018} extend beyond the observing frequency band, showing no sign of a spectral cutoff. This implies that the inferred magnetic field strengths of these T dwarfs (at kG level) are in fact lower limits.
\citet{kao2018} also found that the emission is much more intermittent at higher frequencies and suggested that this is because the emission originates in transient small-scale magnetic structures.
Meanwhile, the LOFAR-band emission from J1019+65 appears to be ever-present given our detection in the inverse-variance-weighted average images. This persistent emission is consistent with the LOFAR-band emission occurring in the large-scale dipolar field of the emitter, as opposed to the centimetre-waveband emission, which may originate in transient small-scale structures. We note that the dynamo scaling laws predict the average field strength in the dynamo region, which is likely more representative of the large-scale field. As such, low-frequency radio observations are important for building up an empirical scaling law for the magnetic field strengths of brown dwarfs and perhaps exoplanets.

We note that while \citet{kao2018} did not find a spectral cutoff for their three T dwarfs, \citet[their Figs.~1 and 2]{rose2023} found a declining spectrum above $\approx 2$\,GHz. However, their decline is markedly more gradual; the Stokes V flux density of their brown dwarf is around $-2$\,mJy to $-4$\,mJy between 0.9 and 1.5\,GHz, and gradually drops to the noise level of about $\pm1$\,mJy at around 2.0\,GHz. The flux density of J1019+65, on the other hand, drops by a factor of about 5 going from 120\,MHz to 168\,MHz, and by a factor of 10 between the LOFAR band (144\,MHz) and the uGMRT band-3 (400\,MHz). For this reason, we suggest that the drop in flux density seen by \citet{rose2023} is likely to be due to spectral evolution rather than a surface cutoff, and we have treated their radio detection as a lower limit on the magnetic field strength.

Figure~\ref{fig:scaling_laws_plot} shows the comparison of the magnetic energy densities of fully convective objects (including J1019+65) to values predicted by different scaling laws. The uncertainty in the magnetic energy density (y-axis) of the data points is due to the factor that relates the dipole magnetic field and the internal magnetic field. We assumed $B/B_{\rm dip}=7^{+8}_{-3}$ \citep{christensen2009}. Meanwhile, the uncertainty in the x-axis of the data points is mainly due to the mass uncertainty in $\rho$. For example, the two T dwarfs in the binary of J1019+65 have masses of $41.43 \pm 18.03\,M_J$ and $32.49 \pm 15.83\,M_J$ \citep{harish_j1019}.

Except for rule \#3, for which J1019+65 is consistent with the prediction of the scaling law, all other scaling laws over-predict the magnetic energy density of J1019+65 by a few orders of magnitude. In contrast, every scaling law also under-predicts the magnetic energy density of the T dwarfs detected by \citet{kao2018} by at least a few orders of magnitude (since their data points are lower limits). The scaling laws also under-predict the limit set on the T8 dwarf by \citet{rose2023}.

What is the source of the discrepancy between our sub-GHz cutoff detection and the non-detections of a cutoff by \citet{kao2018}? One possibility is that we are seeing emission from different magnetic structures. \citet{kao2018} and \citet{rose2023} could be detecting emission from more localised structures that have a higher (spatial) order structure -- quadrupole, octupole, etc. -- whereas we are detecting emission from charges trapped mainly in a large-scale dipole structure. 
Interestingly, recent numerical dynamo simulations on large-scale (mean) magnetic field generation in rotating spherical dynamos by \citet{orvedahl2021} have shown that the large-scale and small-scale magnetic fields saturate via distinct mechanisms. They find that in rapidly rotating dynamos, the large-scale axisymmetric component of the magnetic field follows the Elsasser number rule, consistent with what we saw in rule \#3 in Fig.~\ref{fig:scaling_laws_plot} if we assume that the metre-wave emission originates in the stable large-scale dipolar field (mean field) and the cm-wave emission originates in small-scale transient magnetic structures (fluctuating dynamo).  

Another possibility is that the different frequency ranges are detecting different components of emission, each with their own L-shell locations and ECMI beaming geometries. This is the case for Jupiter's broadband kilometric, hectometric, and decametric (Io and non-Io related) components that each have their own spectral shapes \citep{zarka2004b}. The limited sensitivity of any observation could then make a natural spectral evolution appear as a cutoff. Because the radio spectral shape cannot be predicted from first principles, we cannot rule out this possibility. But we consider it less likely due to the abruptness of J1019+65's cutoff as compared to the typical spectral shape seen in Jovian emission components. For instance, the Jovian hectometric component declines in flux density by an order of magnitude over roughly an order of magnitude in frequency (see Fig.~8 of \citealp{zarka2004b}), whereas J1019+65's cutoff leads to a fall of an order of magnitude just over a factor of two in frequency. Nevertheless, monitoring a sample of methane dwarfs from $\sim 100\,{\rm MHz}$ to $\sim 10\,{\rm GHz}$ is necessary to test if there are indeed multiple emission components in these objects or if our interpretation of a surface ECMI cutoff is correct.

\section{Conclusions}
\label{sec:conclusion}

We observed J1019+65 with three different radio telescopes, amounting to more than 70 hours of observations acquired between 2021 and 2025 and spanning a frequency range from 120\,MHz to 2\,GHz. 
The circularly polarised emission of J1019+65 is only seen in our LOFAR-band data and is conspicuously absent at higher frequencies. We interpret this as the coveted spectral cutoff of ECM emission at a frequency that corresponds to the cyclotron frequency at the polar surface of the emitter. This yields an inferred polar surface magnetic field strength of $\approx 9 B_J$ or around 126\,G.

We compared our results for J1019+65 and its magnetic field strength to the different scaling laws in the literature. We found that except for the Elsasser number rule (rule \#3 in Table~\ref{tab:scaling_laws}), all other scaling laws over-predict the inferred field strength of J1019+65 by a few orders of magnitude.

The discrepancy could arise for either of two reasons: (a) the observed cutoff is a result of frequency-dependent beaming rather than a genuine cutoff; or (b) the GHz-regime observations trace emission from small magnetic structures (possibly transient and driven by a fluctuating dynamo component), whereas J1019+65's observed emission originates in the large-scale field, possibly associated with the so-called mean-field dynamo.

Because the cutoff is located around 200--300\,MHz as seen in Fig.~\ref{fig:j1019-spectrum}, GMRT band-2 (\numrange[range-phrase = --]{125}{250}{MHz}) observations could lead to a direct detection of the abrupt flux cutoff and a test of whether geometric beaming could masquerade as a cutoff. Ultimately, the cutoff locations of many more low-frequency-detected brown dwarfs need to be measured in order to discern any trends with fundamental parameters such as mass, density, rotation rate and thermal luminosity. Our work, especially Fig.~\ref{fig:scaling_laws_plot}, is a step in this direction.

\begin{acknowledgements}
T.W.H.Y. and H.K.V. acknowledge funding from EOSC Future (Grant Agreement no. 101017536) projects funded by the European Union's Horizon 2020 research and innovation programme. HKV acknowledges funding from the European Research Council via the starting grant `STORMCHASER' (grant number 101042416) and from the Dutch research council under the talent programme (Vidi grant VI.Vidi.203.093). JRC acknowledges funding from the European Union via the European Research Council (ERC) grant Epaphus (project number 101166008). LOFAR is the Low Frequency Array designed and constructed by ASTRON. It has observing, data processing, and data storage facilities in several countries, which are owned by various parties (each with their own funding sources), and which are collectively operated by the ILT foundation under a joint scientific policy. The ILT resources have benefited from the following recent major funding sources: CNRS-INSU, Observatoire de Paris and Universit\'e d'Orl\'eans, France; BMBF, MIWF-NRW, MPG, Germany; Science Foundation Ireland (SFI), Department of Business, Enterprise and Innovation (DBEI), Ireland; NWO, The Netherlands; The Science and Technology Facilities Council, UK; Ministry of Science and Higher Education, Poland; The Istituto Nazionale di Astrofisica (INAF), Italy. 
This publication is part of the project LOFAR Data Valorization (LDV) [project numbers 2020.031, 2022.033, and 2024.047] of the research programme Computing Time on National Computer Facilities using SPIDER that is (co-)funded by the Dutch Research Council (NWO), hosted by SURF through the call for proposals of Computing Time on National Computer Facilities.
We thank the staff of the GMRT that made these observations possible. GMRT is run by the National Centre for Radio Astrophysics of the Tata Institute of Fundamental Research. The National Radio Astronomy Observatory and Green Bank Observatory are facilities of the U.S. National Science Foundation operated under cooperative agreement by Associated Universities, Inc.
\end{acknowledgements}

\bibliographystyle{aa}
\bibliography{ref}

\clearpage
\onecolumn
\appendix

\section{Supplementary table}

The details of each follow-up radio observation of J1019+65 are
shown in Table~\ref{tab:observations}.

\setlength{\tabcolsep}{15pt}
\renewcommand{\arraystretch}{1.5}

\begin{table}[H]
    \caption{Long-term radio monitoring of J1019+65 using different radio telescopes.}
    \centering
    \begin{tabular}{c c c c}
        \toprule
        Start of Epoch & Instrument & Frequency band & Exposure time \\
        (UTC) & & & (hours) \\
        \midrule
        2021-02-23T19:01:20.0 & LOFAR & 120 -- 168 MHz & 8   \\
        2022-12-01T03:06:00.0 & LOFAR & 120 -- 168 MHz & 4   \\
        2023-01-15T00:11:00.0 & LOFAR & 120 -- 168 MHz & 4   \\
        2023-01-24T21:41:00.0 & LOFAR & 120 -- 168 MHz & 4   \\
        2023-01-24T22:02:41.8 & GMRT  & 550 -- 950 MHz & 5.5 \\
        2023-02-12T21:41:00.0 & LOFAR & 120 -- 168 MHz & 4   \\
        2023-02-14T22:11:00.0 & LOFAR & 120 -- 168 MHz & 4   \\
        2023-03-05T20:11:00.0 & LOFAR & 120 -- 168 MHz & 4   \\
        2023-03-08T17:35:38.9 & GMRT  & 300 -- 500 MHz & 5   \\
        2023-03-08T18:11:00.0 & LOFAR & 120 -- 168 MHz & 4   \\
        2023-03-11T23:11:00.0 & LOFAR & 120 -- 168 MHz & 4   \\
        2023-03-20T20:11:00.0 & LOFAR & 120 -- 168 MHz & 4   \\
        2023-04-22T21:38:33.0 & VLA   & 1 -- 2 GHz     & 4.5 \\
        2025-01-06T00:47:57.8 & GMRT  & 300 -- 500 MHz & 3.6 \\
        2025-01-07T16:47:31.7 & GMRT  & 300 -- 500 MHz & 7.7 \\
        \bottomrule
    \end{tabular}

    \tablefoot{
        The date and time formats follow ISO 8601.
        The exposure time includes overheads.
    }

    \label{tab:observations}
\end{table}

\end{document}